\documentclass[fleqn,10pt]{wlscirep}
\usepackage{gensymb}

\title{Electron spin relaxations of phosphorus donors in bulk silicon under large electric field}

\author[1,2,+]{Daniel K. Park}
\author[1]{Sejun Park}
\author[1,3]{Hyejung Jee}
\author[1,*]{Soonchil Lee}
\affil[1]{Department of Physics, Korea Advanced Institute of Science and Technology, Daejeon, 34141, Korea}
\affil[2]{Current address: School of Electrical Engineering, Korea Advanced Institute of Science and Technology, Daejeon, 34141, Korea}
\affil[3]{Current address: Department of Physics, Imperial College London, London, SW7 2BW, United Kingdom}

\affil[+]{kpark10@kaist.ac.kr}

\affil[*]{soonchillee@kaist.ac.kr}

\begin{abstract}
Modulation of donor electron wavefunction via electric fields is vital to quantum computing architectures based on donor spins in silicon. For practical and scalable applications, the donor-based qubits must retain sufficiently long coherence times in any realistic experimental conditions. Here, we present pulsed electron spin resonance studies on the longitudinal ($T_1$) and transverse ($T_2$) relaxation times of phosphorus donors in bulk silicon with various electric field strengths up to near avalanche breakdown in high magnetic fields of about 1.2 T and low temperatures of about 8 K. We find that the $T_1$ relaxation time is significantly reduced under large electric fields due to electric current, and $T_2$ is affected as the $T_1$ process can dominate decoherence. Furthermore, we show that the magnetoresistance effect in silicon can be exploited as a means to combat the reduction in the coherence times. While qubit coherence times must be much longer than quantum gate times, electrically accelerated $T_1$ can be found useful when qubit state initialization relies on thermal equilibration.
\end{abstract}
\begin{document}
\flushbottom
\maketitle

\thispagestyle{empty}
\def\ef{V$/\mu$m}
\def\cc{P/cm$^3$}
\section*{Introduction}
Phosphorus donor spins in silicon (Si:P) are promising candidates for encoding quantum information due to outstanding coherence times and the availability of the mature semiconductor industry. Since the quantum computing architecture based on donor spins in silicon was proposed by Kane~\cite{kane}, many significant milestones, such as extending qubit coherence times via silicon-28 isotope enrichment~\cite{PhysRevB.68.193207,0953-8984-18-21-S06,NatureNanotech30s,itoh_watanabe_2014,nmat3182}, high-fidelity control and readout of single donor spins~\cite{SingleDonorDevice1,SingleDonorDevice2,SingleDonorHighFidelity,0953-8984-27-15-154205}, and resonance frequency tuning using electric fields induced Stark shift for the qubit-selective control~\cite{PhysRevLett.113.157601,Lauchte1500022}, have been achieved. Also, several device designs within the donor-based framework have been proposed~\cite{Hille1500707,SiliconSurfaceCode,PhysRevB.93.035306} to enable topological quantum error correction. However, realizing two-qubit quantum operations using the exchange interactions as originally proposed by Kane in a scalable manner remains challenging due to the extreme sensitivity of the interaction strength to the donor displacement. To relax the requirement on the spatial precision, an architecture that exploits electric dipole interactions was recently proposed~\cite{SiFFqubit}. In this approach, the qubit is defined using the flip-flop energy splitting of the nuclear and electron spin states, and is called the flip-flop qubit. One and two qubit gate implementations require shifting the donor electron wavefunction to the ionization point, where the electron is shared halfway between donor and Si/SiO$_2$ interface, via electric fields. This scheme allows for the larger inter-qubit spacing than the exchange-based method, which yields sufficient room to place classical control and readout components~\cite{SiFFqubit}. After all, the application of electric fields is ubiquitous in various proposals for donor-based quantum information processing. On the other hand, theoretical analyses of the single-donor flip-flop qubit with reasonable experimental parameters predicted considerable decrease of the qubit $T_1$ relaxation time due to strong interaction with phonon-induced deformation potentials and the nontrivial valley-related characteristics of the electron-phonon interaction and the involved electronic states~\cite{ValleyEnhancedT1}. Yet, the effect of electric fields near ionization on the $T_1$ and $T_2$ times of donor spins in silicon with higher qubit densities is not clear-cut.

This work reports a detailed experimental study on the spin relaxation times of phosphorus ($^{31}$P) donor electrons in bulk silicon under electric fields ($E_0$) ranging up to near avalanche breakdown triggered by impact ionization using pulsed electron spin resonance (ESR). Two Si:P wafers with donor concentrations of about $10^{14}$-$10^{15}$ \cc\;are tested in the external static magnetic field ($B_0$) strength of about 1.2 T and at temperatures near 8 K for good compromise between the detection sensitivity and the $T_1$ relaxation time in the absence of electric field. The key findings are summarized as follows. The $T_1$ relaxation time changes dramatically under strong electric field near (but before) the avalanche breakdown, and the coherence time $T_2$ is limited by $T_1$. This is attributed to the electric current (I) in the bulk silicon. However, due to magnetoresistance in silicon~\cite{doi:10.1063/1.3238361,NatureMR}, the amount of electric current varies with the orientation of the electric field with respect to the magnetic field. Therefore, the reduction in the relaxation times can be minimized by carefully determining the orientation.

\section*{Results}
\begin{figure}[t]
\centering
\includegraphics[width=0.25\linewidth]{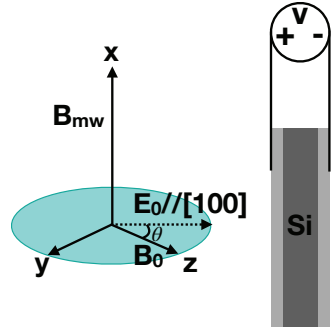}
\caption{Schematic of the Si:P wafer (dark gray) with aluminum (light gray) sputtered on two faces of the wafer (not in scale) and its orientation with respect to external fields. The electric field, $E_0$, is formed between the metal plates by applying DC voltage. The sample is inserted in the static magnetic field, $B_0$, in such a position that the direction of the electric field is coplanar to the magnetic field. The microwave field, $B_\text{mw}$, for controlling electron spins points perpendicular to both $E_0$ and $B_0$. The coordinate frame on the left is arbitrarily defined for convenience.}
\label{fig:1}
\end{figure}

The spin relaxation times of the donor-bound electrons are measured from two Si:P wafers, A and B, with phosphorus concentrations of $2.2\times10^{14}$-$4.9\times 10^{15}$ P/cm$^{3}$ and $3.5\times 10^{14}$-$6.5\times 10^{14}$ P/cm$^{3}$, respectively, using a Q-band ESR spectrometer (see Methods for details). These concentrations correspond to about 59 to 166 nm inter-donor distance assuming uniform distribution. The electric field is formed along the [100] crystal orientation by applying voltage between two 50-nm-thick aluminum plates sputtered on each faces of the wafer. The direction of $E_0$ is coplanar to the external static magnetic field, and perpendicular to the oscillating microwave field ($B_{mw}$) as shown in Figure~\ref{fig:1}.

\begin{figure}[t]
\centering
\includegraphics[width=0.8\linewidth]{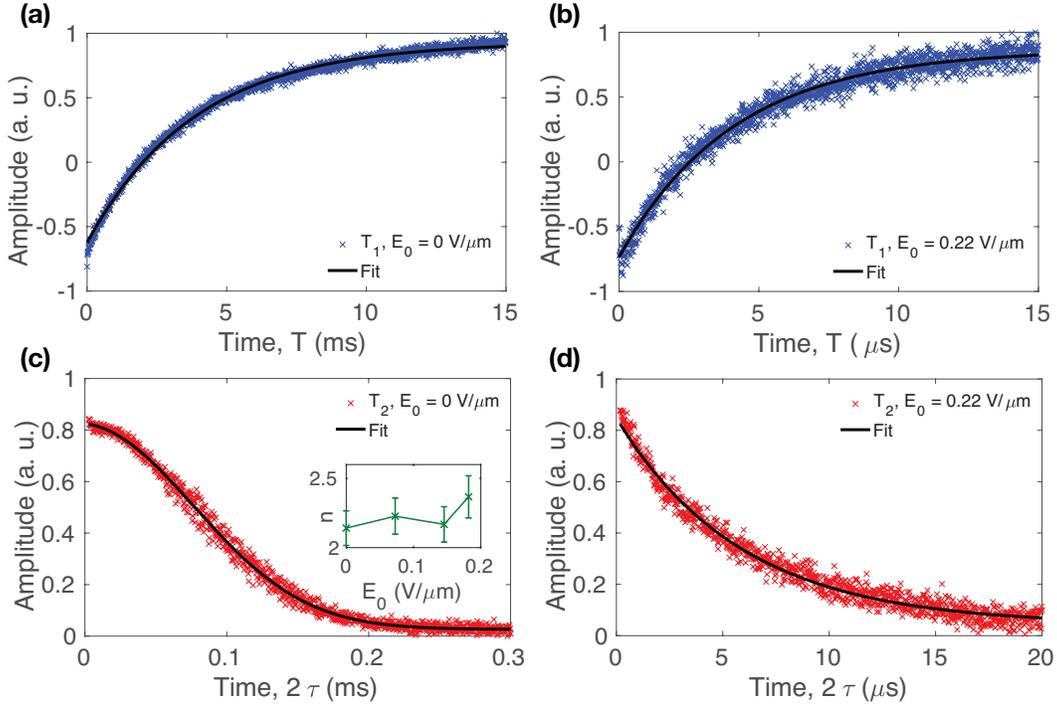}
\caption{Examples of the electron spin echo decay to measure $T_1$ (a) without electric field, and (b) with $E_0=0.22$ \ef, and to measure $T_2$ (c) without electric field, and (d) with $E_0=0.22$ \ef\;for sample A at around 1.2 T and 8 K. The solid line in (c) is the fitting to $s(2\tau)=\exp\lbrack-(2\tau/T_{2a})^n-2\tau/T_{2b}\rbrack$, and the inset shows the fitted values of $n$ with the error bars representing the fitting error. The data in (a), (b), and (d) are fitted using a single exponential decay curve.}
\label{fig:2}
\end{figure}
For both samples, the electron spin echo signal decays in the $T_1$ measurement experiments are single exponential regardless of the magnitude of the electric field. On the other hand, when $E_0=0$, the $T_2$ decay curve is better described by $s(2\tau)=\exp\lbrack-(2\tau/T_{2a})^n-2\tau/T_{2b}\rbrack$, where $s(2\tau)$ is the normalized electron spin echo signal with the interpulse delay $\tau$~\cite{PhysRevB.82.121201}. However, we found that as the $E_0$ value reaches certain regime, the $T_2$ decay becomes a single exponential. In our measurements, the $E_0$ values from which the coherence decay curves are single exponential are $0.22$ \ef\;and $0.13$ \ef\;for sample A and B, respectively. Moreover, at these points, the $T_1$ relaxation times are significantly reduced. Figure~\ref{fig:2} shows examples of electron spin signal decay from the $T_1$ [(a) and (b)] and $T_2$ [(c) and (d)] measurements for sample A with $E_0=0$ and $0.22$ \ef\;at $B_0=1.2$ T and 8 K. The inset in Figure~\ref{fig:2} (c) shows that $n$ estimated from fitting for several values of $E_0$ in the non-single exponential regime ranges between two to three, agreeing with previously reported values attributed to nuclear-induced spectral diffusion~\cite{PhysRevB.82.121201,0953-8984-18-21-S06}. The large difference between the time scale (horizontal axis) of the figures on the left [(a) and (c)] and on the right [(b) and (d)] demonstrates the substantial increase of the relaxation rates when the electric field is turned on.

\begin{figure}[t]
\centering
\includegraphics[width=0.98\linewidth]{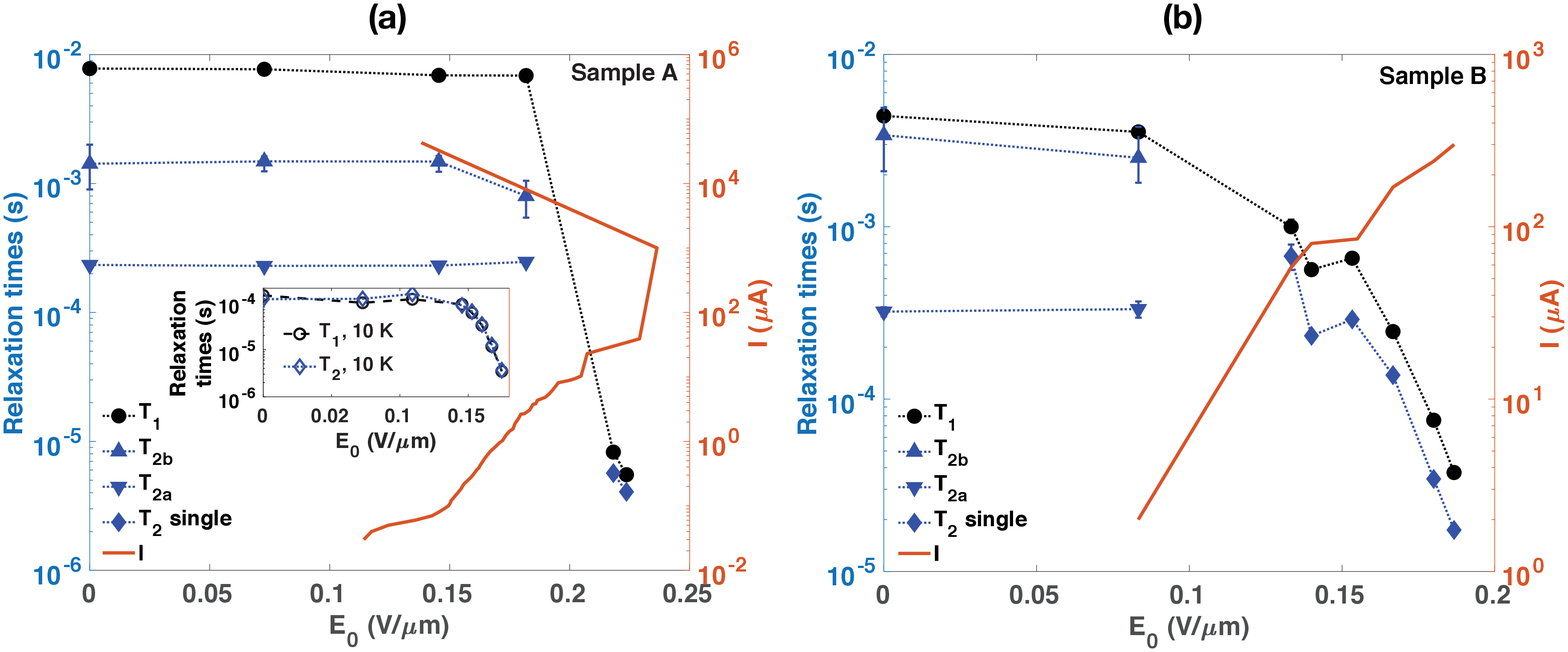}
\caption{The electric field dependence of electron spin relaxation rates (left y-axis) for the Si:P sample (a) A and (b) B at near 1.2 T and 8 K. The inset in (a) shows the relaxation rates as a function of $E_0$ at 10 K. The circles represent the $T_1$ time, two triangle symbols are obtained from the non-exponential signal decay in the $T_2$ measurement, and the diamonds represent the $T_2$ time in the large electric field regime where the signal decay becomes single exponential. The solid lines (right y-axis) represent the electric current between two aluminum plates on the faces of the silicon wafer as a function of $E_0$. In (a), the avalanche breakdown occurs around $0.24$ \ef\; as indicated by the discontinuity in the current-electric field curve. The error bars represent fitting errors, and are smaller than the data symbol when not visible.}
\label{fig:3}
\end{figure}

The relationship between $E_0$ and the relaxation times is plotted in Figure~\ref{fig:3}. For A, $T_1$ is reduced by about three orders of magnitude at $E_0\ge 0.22$ V$/\mu m$. The coherence decays fit well to single exponential in this regime, and $T_2$ appears to be limited by $T_1$. The solid line represents the magnitude of electric current between the metal plates, and shows the avalanche breakdown, a sudden transition to a low-resistance state, occurs at about $0.24$ \ef\;for this sample. Thus, $E_0$ could not be increased beyond this point. Interestingly, the relaxation times change dramatically even though the current is about two orders of magnitude smaller than that at the breakdown. The inset in Figure~\ref{fig:3} (a) shows the relaxation times in A as a function of $E_0$ at 10 K to demonstrate that the effect persists in a different temperature. In this temperature and donor concentration, $T_2$ is limited by $T_1$ as the $T_1$ process dominates decoherence~\cite{nmat3182}. The relaxation times decrease rapidly as $E_0$ is increased beyond $0.15$ \ef, similar to the behaviour observed at 8 K. The electric field dependence of the relaxation times is qualitatively confirmed with the sample B as shown in Figure~\ref{fig:3} (b). Both $T_1$ and $T_2$ do not exhibit noticeable changes until $E_0$ is increased up to about 0.08 \ef. But from beyond this point, $T_1$ undergoes about two orders of magnitude reduction as $E_0$ is increased up to about 0.19 V$/\mu m$. Meanwhile, the $T_2$ decay converges to single exponential, and the spin coherence time appears to be limited by the $T_1$ process in this regime. The experimental data provide clear evidence that the longitudinal relaxation time of electron spins in Si:P is reduced substantially even when the applied electric field is smaller than the breakdown field. On the other hand, the impact of the electric field on $T_2$ independent from the $T_1$ process is uncertain. The electric field effects on the spin relaxation times of the donor-bound electron in bulk silicon implies that the operating conditions of the Si:P quantum devices with higher donor densities can be limited due to accelerated decoherence. Hereinafter, we focus on the effect of the electric field on $T_1$.

\begin{figure}[t]
\centering
\includegraphics[width=0.5\linewidth]{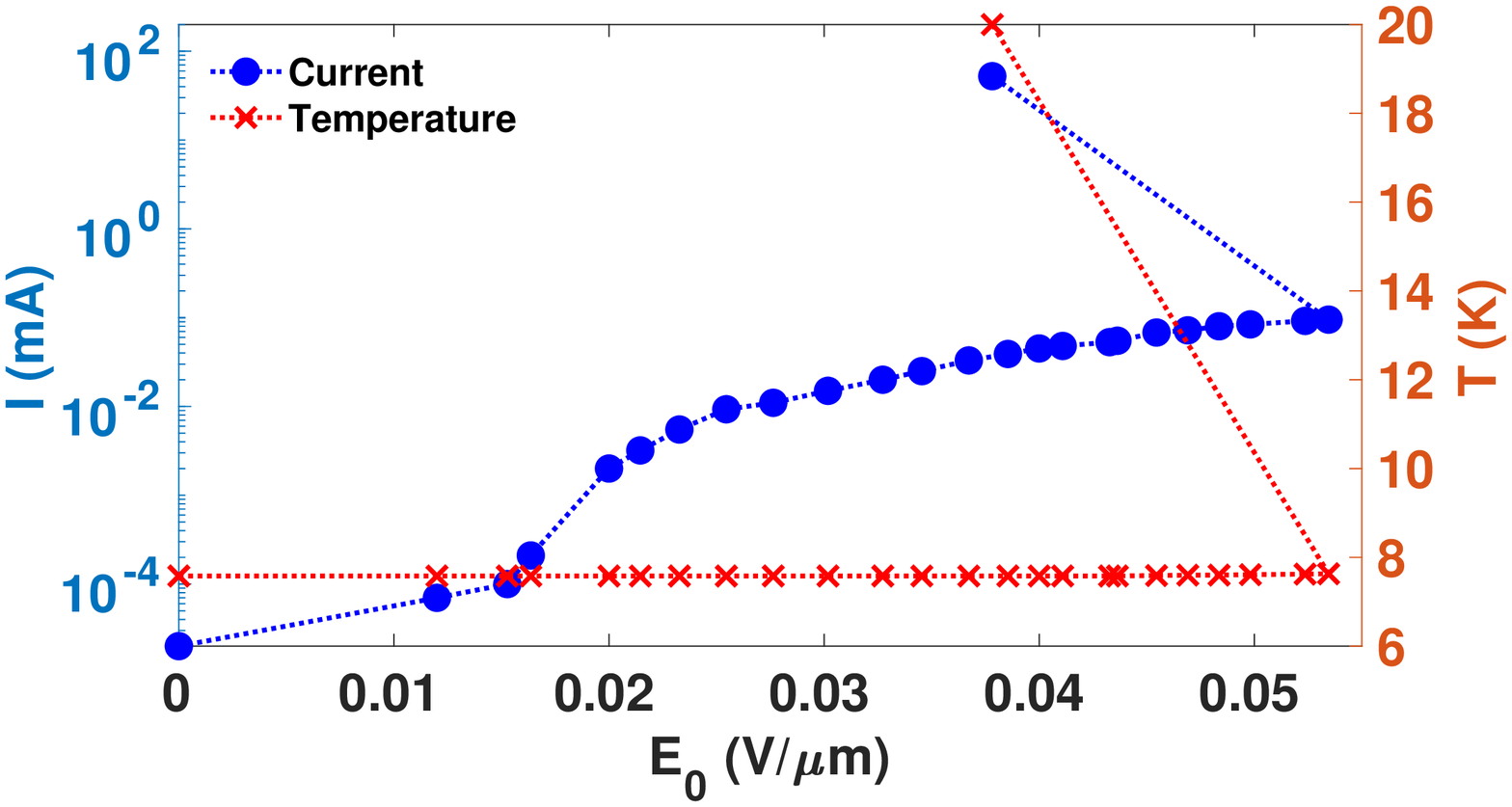}
\caption{The electric current between two aluminum plates on the faces of the silicon sample A (left y-axis) and the sample temperature (right y-axis) as a function of the electric field strength, $E_0$. The avalanche breakdown is observed by the discontinuity in the current-electric field curve.}
\label{fig:4}
\end{figure}
In the temperature range in which the measurements are conducted, the strong temperature dependence of $T_1$ is known to exist due to spin-phonon relaxation process~\cite{PhysRev.114.1245,PhysRev.155.816}. As a first step towards understanding the source of the electric field dependence of the relaxation times, we measured the sample temperature and the electric current in a Si:P wafer between two metal plates with respect to $E_0$ (see Methods). The measurement results depicted in Figure~\ref{fig:4} show that the sample temperature remains nearly constant until the breakdown field, at which the resistivity of the sample abruptly drops, is reached. Thus, the dramatic $T_1$ reduction in our measurements cannot be explained by the change in the sample temperature.

\begin{figure}[t]
\centering
\includegraphics[width=0.98\linewidth]{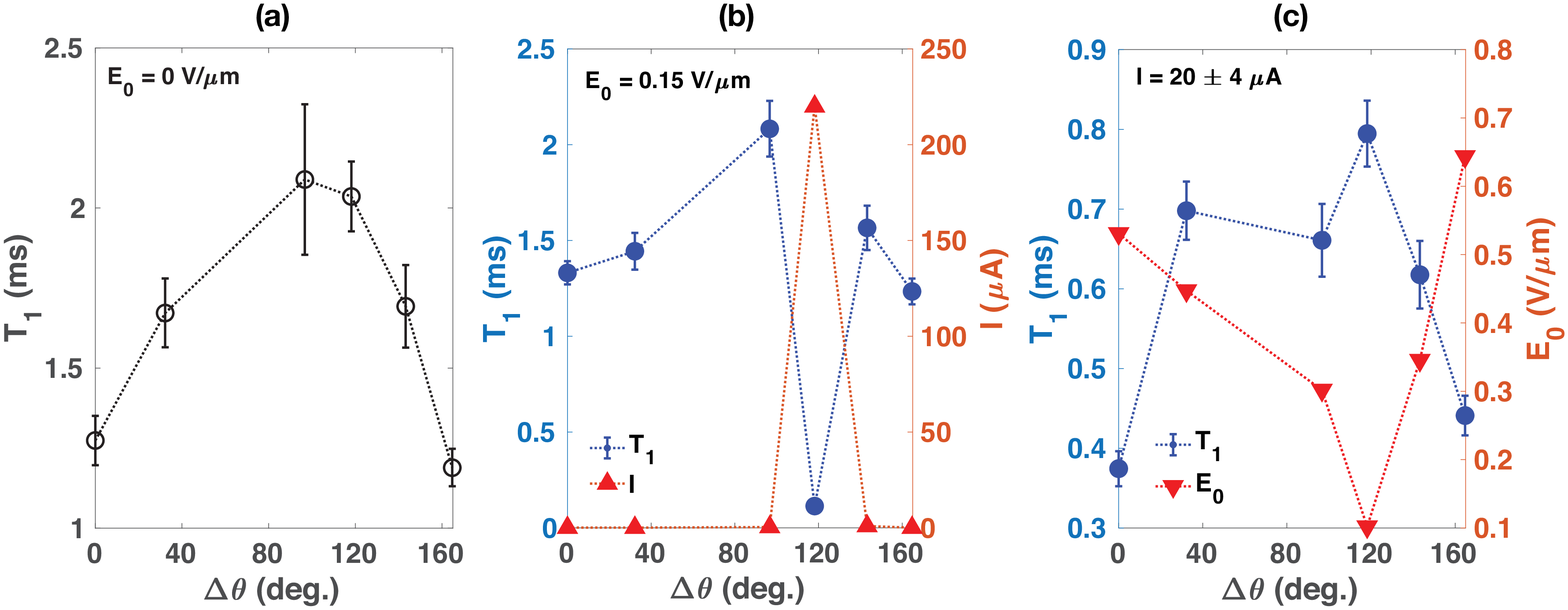}
\caption{\label{fig:5}The $T_1$ relaxation times for the sample B (left y-axis) as a function of an angle between $E_0$ and $B_0$ when (a) $E_0=0$, (b) $E_0=0.15$ \ef, and (c) $I=20\pm 4$ $\mu$A. $\Delta\theta$ corresponds to the change of the angle between $E_0$ and $B_0$. For a fixed value of $E_0$, the electric current varies with the orientation, and vice versa. The anisotropy of the current and $E_0$ are presented in (b) and (c), respectively (right y-axis). The vertical error bars represent fitting errors. The uncertainty in $\Delta\theta$ is smaller than the width of the symbols.}
\end{figure}
Next, we experimentally investigated the anisotropy of $T_1$ in the crystal orientation with respect to $B_0$ using B, the sample with more uniform donor distribution. In this study, the silicon crystal, and hence the direction of $E_0$, were rotated around the x-axis defined in Figure~\ref{fig:1} while the direction of $B_0$ was fixed. For each orientation, we conducted $T_1$ measurements with $E_0=0,\; 0.15$ \ef, and at constant electric current of $20\pm 4$ $\mu A$ measured in the same direction as $E_0$. To maintain the constant electric current at each orientation, the strength of $E_0$ is adjusted by controlling the DC voltage. The results are shown in Figure~\ref{fig:5}. Note that the independent variable in the figure ($\Delta\theta$) corresponds to the change of the angle between $E_0$ and $B_0$ from the reference angle determined by the initial sample placement in the magnetic field. The $T_1$ relaxation time is observed to be anisotropic in the crystal orientation, and hence the direction of $E_0$, with respect to $B_0$ in all experiments. For a fixed amount of $E_0$, we observed that the electric current in [100] direction also varied with respect to the orientation between $E_0$ and $B_0$. As shown in Figure~\ref{fig:5} (b), when $E_0=0.15$ \ef, the current is maximum near $\Delta\theta=120\degree$. Interestingly, $T_1$ at this orientation is about an order of magnitudes smaller than that at other orientations. The anisotropic $T_1$ is also observed when the current is fixed at about $20$ $\mu$A as illustrated in Figure~\ref{fig:5} (c). In this case, however, the qualitative feature of the orientation dependence of $T_1$ resembles that of the orientation dependence observed when the electric field is turned off (Figure~\ref{fig:5} (a)), except the relaxation times at each point are reduced by approximately a factor of three.

\section*{Discussion}
The spin relaxation anisotropy without the electric field displayed in Figure~\ref{fig:5} (a) is consistent with the previous results attributed to a modulation of the electronic $g$ factor by acoustic phonons~\cite{PhysRev.118.1523,PhysRev.118.1534,PhysRev.124.1068}. The previous studies show that the $T_1$ relaxation time is the longest when $B_0$ is aligned with [100] axis of the silicon crystal~\cite{PhysRev.124.1068}. Thus, in our experimental data, the [100] axis lies within the range of $\Delta\theta=96\degree$ to $118\degree$.

Figure~\ref{fig:5} (b) shows that the electric current in the wafer between two metal plates is the largest when $B_0$ is aligned with [100] axis, along which $E_0$ is applied. The variation of the electric current with respect to the angle between $E_0$ and $B_0$ for a fixed electric field strength is consistent with the effect of magnetoresistance (MR), defined as $[R(B)-R(0)]/R(0)$, where $R(B)$ is the resistance in magnetic field $B$. The phosphorus density of sample B is just within the low doping regime where the MR for the perpendicular orientation ($B_0\perp E_0$) is large~\cite{doi:10.1063/1.3536663}. For Si:P, the MR effect persists for the longitudinal orientation, i.e., $B_0\|E_0$, since the Lorentz force can still deflect the carriers as the trajectories are distorted due to the random distribution of donors. Nevertheless, the longitudinal MR is known to be much smaller than the perpendicular MR~\cite{doi:10.1063/1.3536663,C3NR04077A}. Therefore, the large electric current at a specific orientation ($\Delta\theta\approx 118\degree$) can be interpreted as a consequence of the minimum MR since $E_0$ is parallel to $B_0$.

When $E_0=0.15$ V$/\mu$m, although the electric field is uniformly raised for all orientations, the $T_1$ time is significantly reduced only when the electric current is high. On the other hand, Figure~\ref{fig:5} (c) shows that despite the large variation of $E_0$ with respect to the orientation, all relaxation times are about a factor of three smaller than the relaxation times at the same orientation without the electric field. These experimental results suggest that the acceleration of the longitudinal relaxation rate at strong electric field is related to the rise of the electric current in the Si:P crystal rather than the strength of the electric field.

The quality of the electron spin qubits in Si:P can be degraded considerably when large electric field is needed, due to the increased decoherence rate. In order to enhance the utility and the flexibility of the donor-based spin qubits, it is desirable to extend the high-fidelity qubit operating range to strong electric fields. Moreover, there exists an architectural proposal that demands the application of strong electric fields so that the electron spins are manipulated near the ionization point~\cite{SiFFqubit}. Therefore, finding strategies to circumvent the escalation of the spin relaxation rates is critical. From above experimental studies, we found that the electric current, rather than the electric field, is responsible for the rapid change in $T_1$. Then we experimentally verified that MR yields the electric current anisotropy in the orientation of $E_0$ with respect to the external magnetic field. Therefore, if the angle between $E_0$ and $B_0$ is chosen properly, the electric current for a given $E_0$ strength, and hence the reduction in the $T_1$ time can be minimized. In particular, the electric field should not be aligned with the magnetic field. For instance, Figures~\ref{fig:5} (b) and (c) illustrate that when $E_0$ is nearly parallel to $B_0$ ($\Delta\theta\approx 120\degree$), $T_1$ is about $0.1$ ms for $E_0=0.15$ V$/\mu$m, while when $E_0$ is nearly perpendicular to $B_0$ ($\Delta\theta\approx 40\degree$), $T_1$ is about $0.7$ ms for $E_0=0.45$ V$/\mu$m. Thus, in this example data, the $T_1$ time at one orientation can be about seven times longer than that at another orientation, although about three times larger $E_0$ is used. Recall that in the absence of $E_0$, the $T_1$ time is the longest when $B_0$ is applied along the [100] direction. Thus, we speculate that further $T_1$ optimization is possible by using a silicon wafer grown in a direction such that the direction of $B_0$ can be parallel to [100], but perpendicular to $E_0$.

In summary, the $T_1$ relaxation rate of the phosphorus donor electron spins in bulk silicon with low dopant concentration can be increased significantly under large electric fields due to electric current in the sample. The coherence time is also shortened as it is upper-bounded by $T_1$. On the other hand, the amount of electric current is anisotropic in the $E_0$ orientation with respect to the external magnetic field due to the MR effect. Thus, the reduction in the relaxation rates for a fixed electric field strength can be minimized by choosing the appropriate orientation. Although the decoherence rate must be minimized during quantum gate operations, the fast $T_1$ relaxation can be exploited in a special instance. Namely, when the qubit initialization method relies on thermal equilibration, the fast relaxation rate is favoured for resetting the qubits. Furthermore, the ability to engineer the $T_1$ relaxation rate can be useful for dynamic nuclear polarization since the increased $T_1$ rate allows for the faster polarization of nuclear spins at the cost of the higher microwave pulse power for saturating the electron spin transitions~\cite{PhysRevB.90.214401}. Future work could extend the range of the experimental conditions, such as the magnetic field strength, the temperature, and the donor density. Our results also motivate further studies on the nuclear spin relaxation rates, as well as the case of the single donor spins in large electric fields.

\section*{Methods}
\subsection*{Sample preparation}
Two commercially purchased phosphorus doped (100)-silicon wafers with natural abundance (4.7$\%$ of $^{29}$Si) are used throughout the experiments. The first wafer (A) is quoted with the room temperature resistivity of 1-20 $\Omega\cdot\text{cm}$ (about $2.2\times10^{14}$-$4.9\times 10^{15}$ P/cm$^{3}$), and the thickness of $275$ $\mu$m. The second wafer (B) has the room temperature resistivity of 7-13 $\Omega\cdot\text{cm}$ (about $3.5\times 10^{14}$-$6.5\times 10^{14}$ P/cm$^{3}$), and is $300$  $\mu$m thick. The wafers are cut to a size of approximately 1 mm $\times$ 15 mm to fit in a standard Q-band ESR tube. 
\subsection*{Electron spin resonance measurements}
All ESR experiments were carried out at Korea Basic Science Institute (KBSI) in Seoul, Korea. 34 GHz Q-band pulsed ESR data were obtained on a Bruker Elexsys E580 spectrometer using an EN5107D2 resonator. Cryogenic temperatures were achieved with an Oxford CF-935 cryostat and an Oxford ITC temperature controller.
\subsection*{Relaxation times measurements}
$T_1$ is measured via inversion recovery experiment, and $T_2$ is measured via Hahn echo decay experiment. The pulse sequence for the inversion recovery experiment can be expressed as $\pi-T-\pi/2-\tau-\pi-\tau-\text{echo detection}$. The delay, $T$, after the first $\pi$ pulse was varied while $\tau$ was fixed, and the amplitude of the primary echo signal formed by the second and third pulses was measured. The pulse sequence for the Hahn echo decay experiment is $\pi/2-\tau-\pi-\tau-\text{echo detection}$, and the amplitude of the echo signal was measured as a function of the delay, $\tau$. The $\pi/2$ and $\pi$ pulse lengths were 16 and 32 ns, respectively, in both experiments.
\subsection*{Current-voltage measurements}
The current-voltage relations are measured using two multimeters (Fluke 287 True-RMS), each connected in parallel and in series with the silicon wafer for measuring voltage and current, respectively.
\subsection*{Sample temperature measurements}
The sample temperature dependence on the electric field strength was measured with an Si:P piece cut from A to an approximate size of 1 cm${^2}$. A calibrated temperature sensor (Lakeshore DT-470-CU-13) was attached on top of the aluminum on one face of the wafer, and the other face of the wafer was in contact with the copper heat-sink of a low temperature probe. Then the probe was cooled using a helium-flow cryostat. The current was measured using a multimeter while external DC voltage was applied.
\subsection*{Data availability}
The datasets generated during and/or analysed during the current study are available from the corresponding author on reasonable request.

\section*{Acknowledgements}
We thank Sungmin Kwon and Sumin Lim for helpful discussions, and Donghyuk Jeong and Yujeong Kim for operating the ESR spectrometer at KBSI. This research was supported by the National Research Foundation of Korea (Grants No. 2015K1A31A14021146, 2015R1A2A2A01006251, and 2016R1A5A1008184).
%
\end{document}